\documentclass[a4paper,pre,twocolumn,superscriptaddress]{revtex4}
\usepackage[english]{babel}
\usepackage{inputenc}
\usepackage{graphicx,epsfig,placeins}
\usepackage{latexsym}
\usepackage{hyperref}
\usepackage{amsfonts} 
\usepackage{amsmath,amssymb,amsfonts}
\usepackage{textcomp}
\usepackage{epstopdf}
\usepackage{color}
\usepackage{appendix}

\def\ket|#1>{| #1 \rangle}
\def\bra<#1|{\langle #1 |}
\def\<{\left\langle}
\def\>{\right\rangle}
\def\({\left(}
\def\){\right)}
\def\[{\left[}
\def\]{\right]}
\def\{{\left\lbrace}
\def\}{\right\rbrace}
\def\beq{\begin{equation}}
\def\eeq{\end{equation}}

\def\N{{\mathbb N}}
\def\R{{\mathbb R}}

\def\eff{{\text{eff}}}

\begin{document}

\title{Understanding the enhanced synchronization of delay-coupled
  networks with fluctuating topology}

\author{Otti D'Huys} 
\affiliation{Department of Mathematics, Aston University, B4 7ET
  Birmingham, United Kingdom}%
  
 \author{Javier Rodr\'iguez-Laguna} 
\affiliation{Departamento de F\'isica Fundamental, UNED, Spain}%

\author{Manuel Jim\'enez}
\affiliation{Departamento de F\'isica Fundamental, UNED, Spain}%

\author{Elka Korutcheva}
\affiliation{Departamento de F\'isica Fundamental, UNED, Spain}%
\affiliation{G. Nadjakov Inst. Solid State Physics, Bulgarian Academy
  of Sciences, 1784, Sofia, Bulgaria} 

\author{Wolfgang Kinzel}
\affiliation{Institut f\"ur theoretische Physik, Universit\"at
  W\"urzburg, Am Hubland, 97072 W\"urzburg}

\begin{abstract}
We study the dynamics of networks with coupling delay, from which the connectivity changes over time. The synchronization properties are shown to depend on the interplay of three time scales: the internal time scale of the dynamics, the coupling delay along the network links and time scale at which the topology changes.  Concentrating on a linearized model, we develop an analytical theory for the stability of a synchronized solution.  In two limit cases the system can be reduced to an ``effective" topology: In the fast switching approximation, when the network fluctuations are much faster than the internal time scale and the coupling delay, the effective network topology is the arithmetic mean over the different topologies. In the slow network limit, when the network fluctuation time scale is equal to the coupling delay, the effective adjacency matrix is the geometric mean over the adjacency matrices of the different topologies. In the intermediate regime the system shows a sensitive dependence on the ratio of time scales, and specific topologies, reproduced as well by numerical simulations. Our results are shown to describe the synchronization properties of fluctuating networks of delay-coupled chaotic maps.
\end{abstract}

\date{\today}

\maketitle

\section{Introduction}

In many interacting systems the transmission time for information exceeds the time scale of the internal node dynamics. Hence, delay-coupled networks are relevant in a variety of fields, including coupled optical or opto-electronic systems, communication and transportation systems, social networks and biological networks as gene regulatory and neural systems. For example, in the brain a coupling delay between interacting neurons arises from the conduction time of an electric signal along the axon \cite{Buzsaki}, while it accounts for the traveling time of light between lasers (see \cite{Heil, Nixon, Soriano} and references therein). In engineering networks delayed interactions are discussed in the context of transport and mobility issues \cite{Szymanski}, power grids control or complex supply networks \cite{Timme1, Timme2}.

One of the most important studied properties of interacting elements is the ability to show synchronized behavior \cite{Pikovsky, Boccaletti1, Boccaletti2, Arenas1}.  The synchronization patterns allowed a network have been shown to relate to the network symmetries \cite{Golubitsky}; in a network of identical elements the stability of a symmetric state can be directly related to the spectral properties of the adjacency matrix by the Master Stability Function \cite{Pecora}. In delay-coupled networks this connection between (zero-lag) synchronization and spectral properties of the coupling matrix is even simpler \cite{Heiligenthal, Flunkert}; stability is shown to depend on the magnitude of the spectrum in a monotonous way. 

These results are restricted to network connections that are constant in time. However, it is often more realistic to consider a coupling topology that fluctuates. Such time-varying systems arise in a broad range of systems such as (and not limiting to) moving agents, social networks and synaptic plasticity in neural networks \cite{AB, Irribaren, Cattuto, Karsai, Tsukada, Tsuda, Buonomano, Holme}. Synchronization in such time-varying networks is being studied in the context of diffusive coupling of moving oscillators \cite{Peruani, Majhi,Beardo}, chaotic units \cite{Frasca, Fujiwara} and genetic oscillators moving on lattices \cite{Uriu1, Uriu2}, while consensus problems have been investigated in small-world networks of agents with switching topology and time-delay, relying on algebraic graph theory, random matrix theory and control theory \cite{Olfati-Saber2, Liu}. In the context of neural networks with delay, synchronization transitions induced by the fluctuation of adaptive strength were recently reported \cite{Wang}. Similar results have been found in the case of developing neural networks \cite{Chao} or spike-timing dependent plasticity \cite{Knoblauch}. 

A common result in all these problems is the so-called ``fast switching approximation'' \cite{Fast}: if the network topology changes faster than the internal node dynamics, the system can be approximated by a constant topology, that is the {\em arithmetic} mean of the topology over time. However, a full understanding of the dynamics, if the network time scale and internal time scale interfere, is still lacking, to the best our knowledge.

In a previous publication \cite{Javi} we numerically studied synchronization properties of delay-coupled networks with a time-varying topology. We considered an interaction network of coupled chaotic maps with a single coupling delay $\tau$, with a topology fluctuating among an ensemble of small-world networks, with a characteristic time-scale $T_n$. We found that random network switching may enhance the stability of synchronized states, depending on the interplay between the time-scale of the delayed interactions $\tau$ and that of the network fluctuations $T_n$. If the network switching is fast $T_n\ll\tau$, a strong enhancement of the synchronizability of the network has been observed, in the sense that synchronization is stabilized compared to a typical network of the ensemble. This result is in qualitative agreement with the fast switching approximation \cite{Fast}, although the network time exceeds the internal time scale of the nodes.

Here, in order to understand the physics behind the results obtained in \cite{Javi}, we develop an analytical theory in the linearized limit, based on the Master Stability Function. We express the ``effective" connectivity as a function of the three time scales: the internal time scale $T_{in}$, the characteristic time for network fluctuations $T_n$ and the interaction delay time $\tau$. Three cases are investigated: When the network fluctuations are much faster than the internal time scale and the coupling delay $(T_n \ll T_{in}, \tau)$, the effective network adjacency matrix is the {\em arithmetic} average over the different adjacency matrices, as in the fast switching approximation. When coupling delay and network fluctuation time scales are equal $(T_{in} \ll T_n = \tau)$, in the slow network approximation, the effective adjacency matrix is the {\em geometric} mean over the different adjacency matrices. Thirdly, if all three time scales are separated $T_{in}\ll T_n\ll\tau$, we show that the dynamics depends sensitively on the ratio of time scales and the properties of the temporal topologies.

The paper is organized as follows: In Section II we present the modelling equations and apply the Master Stability Formalism to the corresponding linearized model.  In Sections III and IV we discuss the behavior of the system in the fast and slow network approximation, respectively. The interplay of time scales is largely discussed in Section V by paying attention to the parity effect in the relation between $\tau$ and $T_n$. Sections VI and VII present our numerical simulations both for linear and nonlinear systems, respectively. The last Section is devoted to conclusions. The details of the more elaborate analytical calculations are presented in the Appendices.


\section{Modelling equations}
\label{sec:model}

We start from a general model of $N$ identical scalar elements coupled with interaction delays,

\begin{equation}
  \dot{x}_i(t) = f(x_i) +
  \sum_{j}^N A_{ij}(t) g\left({x}_j(t-\tau)\right)\label{eq:model}\,,
\end{equation}
with $x_i \in\R$. The coupling topology is modelled by a time-varying
$N\times N$ adjacency matrix $A(t)$, whose rows add up to one to
ensure the existence of a permutation symmetric state. The coupling
delay $\tau$ is constant over the links.

To determine the stability of a symmetric solution
$x_1(t)=x_2(t)=\hdots, x_N(t)\equiv x(t)$, (i.e a symmetric fixed
point, an in-phase oscillatory solution or a chaotic state in complete
synchronization), the modelling equation is linearized,

\begin{equation}
  \dot{x}_i (t) = f'(x_i(t)) x_i +
  \sum_{j}^N A_{ij}(t) g'(x_j(t-\tau)) x_j(t-\tau)\,,
\end{equation}
where $f'$ and $g'$ are the derivatives of the functions $f$ and $g$
respectively, evaluated along the symmetric solution ${x}(t)$.

We consider the simplest case, with constant coefficients: The first
term $f'(x(t))\equiv -\lambda_0$ represents an ``internal'' decay rate of the nodes; if the nodes are chaotic, it reduces to the (opposite) sub-Lyapunov exponent \cite{Heiligenthal}. The term $g'(x(t-\tau))\equiv \kappa$ represents the coupling strength. The linearized model can be rewritten as

\begin{equation}
  \dot{\bf x} (t) =
  -\lambda_0 {\bf x} +\kappa A(t) {\bf x}(t-\tau)\label{eq:linmodel}\,,
\end{equation}
where ${\bf x}(t)=(x_1(t),\hdots, x_N(t))^T$. In a constant network it
is straightforward to solve this system analytically by calculating
the Master Stability Function. Evaluating Eq. \eqref{eq:linmodel}
along the eigenvectors $v_k$ of the adjacency matrix $A$, on finds

\begin{equation}
  \dot v_k (t) =
  -\lambda_0 v_k + \kappa\gamma_k \left(v_k(t-\tau)\right)\label{eq:MSF}\,,
\end{equation}
with $\gamma_k$ the eigenvalue associated to the eigenvector
$v_k$. The system evolves then exponentially with a rate given by the
Master stability function $\lambda(\gamma_k)$. In the limit of long
delays $\tau\gg\lambda_0^{-1}$, and in the absence of a strongly unstable solution, $\lambda_0 >0$,
the set of exponential solutions of Eq. \eqref{eq:MSF}
$\{\lambda(\gamma_k)\}$ can be written as a pseudo-continuous spectrum
\cite{Yanchuk}

\begin{equation}
  \lambda(\gamma, \omega)=
  i\omega+\frac{\mu(\omega)}{\tau}=
  i\omega + \frac{1}{\tau}
  \ln\left|\frac{\kappa\gamma}{\lambda_0+i\omega}\right|\,.
  \label{eq:pseudospec}
\end{equation}
This result applies for steady states, or for simple chaotic systems
with a constant slope, as the Bernoulli map. However, it is also a
first order approximation for chaotic systems \cite{Thomas} and reproduces the scaling properties of the spectrum of
Lyapunov exponents of chaotic systems with time-delay $\tau$ and
eigenvalues of magnitude $|\gamma|$ \cite{Heiligenthal,Flunkert}.

Here, we consider a coupling matrix $A(t)$ that is not constant: it
changes discontinuously after a network time $T_n$, running through a
sequence of network topologies as $A_1, A_2, \hdots$. Thus, the system is non-autonomous, with time-dependent parameters \cite{Stefanovska}. However, in the following we show that, in certain limits, the synchronization properties of the system under a fluctuating topology can be described with a constant ``effective'' coupling topology $A_\eff$, allowing for to calculate a master stability function for nodes coupled with a time-varying topology.


\section{Fast network approximation}
\label{sec:fast}

In instantaneously coupled networks, it is well known that, if the
network changes fast enough, the effective network is the average
network over time \cite{Fast}. This so-called ``fast switching
approximation'' is valid as well in delay-coupled networks. Indeed, if
$T_n\ll \lambda_0^{-1}$, one can approximate

\begin{eqnarray}
{\bf x} (t_0+T_n)& \approx & {\bf x}(t_0)+T_n \dot{\bf x}(t)\\
& \approx &
{\bf x}(t_0)+T_n \left(-\lambda_0 {\bf x}(t_0) +
\kappa A_1{\bf x}(t_0-\tau)\right),
\nonumber
\end{eqnarray}
so that at $t=t_0+MT_n$, up to first order in $T_n$, we retrieve

\begin{eqnarray}
  \dot{\bf x}(t_0) & \approx &
  \frac{1}{MT_n} \left({\bf x}(t_0+MT_n)-{\bf x}(t_0)\right)\nonumber\\
  & \approx &
  -\lambda_0 {\bf x}(t_0) + \frac{\kappa}{M}
  \displaystyle\sum_{m=1}^M A_m{\bf x}(t_0-\tau)\,.
\end{eqnarray}
This leads to an ``effective'' adjacency matrix

\begin{equation}
A_\eff=\frac{1}{M}\displaystyle\sum_{m=1}^M A_m\label{eq:fast}\,.
\end{equation}

We illustrate this result for a simple network of three coupled nodes, that alternates regularly between two topologies. The adjacency matrices $A_1$ and $A_2$ are given by
\beq
A_1=\begin{pmatrix} 2/3 & 1/3 & 0 \\ 0 & 2/3 & 1/3 \\ 1/3 & 0 & 2/3
\end{pmatrix},\quad
A_2=\begin{pmatrix} 0 & 0 & 1 \\ 1 & 0 & 0 \\ 0 & 1 & 0
\end{pmatrix}.
\label{eq:comm_choice}
\eeq
According to the fast switching approximation, the nodes evolve
exponentially. The decay or growth rate is given by the most unstable
solution of Eq. \eqref{eq:pseudospec}, using the effective adjacency
matrix $A_\eff$ given by Eq. \eqref{eq:fast}. In this particular case,
both adjacency matrices commute. Therefore, the eigenvalues
$\gamma_\eff$ of the effective network are simply the arithmetic
average of the eigenvalues of both networks, along the same
eigenvectors. For the exponential decay rate, we find

\begin{equation}
  \lambda \approx
  \frac{1}{\tau}
  \ln\left|\frac{\kappa\gamma_\eff}{\lambda_0}\right|\label{eq:trans_lam}\,.
\end{equation}
Along the synchronization manifold, we have $\gamma_\eff=1$ by
construction, and hence the dynamics on the synchronization manifold
is the same as in a constant network. The tranverse stability, which
determines the synchronization properties of the network, is
determined by the second largest eigenvalue

$$\gamma_\eff=\frac{\gamma_1+\gamma_2}{2}\,,$$
with $\gamma_1$ and $\gamma_2$ the respective eigenvalues of $A_1$ and
$A_2$ along one of the transverse eigenvectors (in this particular
case both transverse eigenvalue are complex conjugates, hence both
transverse directions are equally stable). The transverse evolution is
measured by the variance over the nodes. We have

\beq
\mu(t)={1\over N}\sum_{i=1}^N x_i(t), \qquad
\sigma^2(t) = {1\over N}\sum_{i=1}^N (x_i(t)-\mu(t))^2.
\eeq
so that the transverse decay rate (TDR), $\lambda$
can be estimated from the evolution of the
variance as

\beq
\sigma^2(t) \sim \exp(2 \lambda t).
\label{eq:def_SLE}
\eeq

\begin{figure}
\includegraphics[width=8cm]{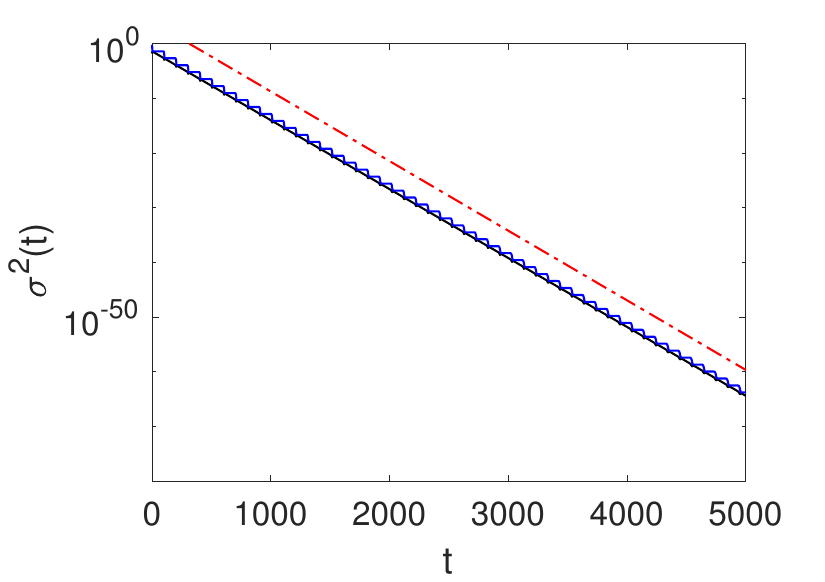}
\caption{Evolution of the variance of the linear system
  for a systems topologies alternating between the
  matrices given in Eq. \eqref{eq:comm_choice}. For the full black,
  almost straight line, initial conditions were exponentially
  decaying, ${\bf x}(t<0)={\bf x}_0 e^{-\lambda t}$, with $\lambda$
  given by the theoretical prediction Eq. \eqref{eq:trans_lam}. For
  the full blue line, with an almost stepwise evolution, we used
  constant initial conditions. The theoretical decay rate (Eq.
  \eqref{eq:trans_lam}) is shown for comparison (red dotdashed line),
  the agreement is excellent. Parameters are $N=3$, $\kappa=0.8$,
  $\lambda_0=1$, $T_n=0.01$ and $\tau=100$.} 
  \label{fig:fastnetworklimit}
\end{figure}

We compare the theoretical decay rate (Eq. \eqref{eq:trans_lam}) with
the numerically calculated evolution of the variance
(Eq. \eqref{eq:def_SLE}) in Fig. \ref{fig:fastnetworklimit} for two
different initial conditions. We find that for the two different
initial conditions that we used, the agreement between theory and
simulations is excellent.

\section{Slow network approximation}
\label{sec:slow}

Let us consider the situation in which the network time $T_n$ is equal
to the coupling delay $\tau$, and both are larger than the
instantaneous decay rate $\lambda_0^{-1}$ of the nodes,
$\lambda_0^{-1}\ll T_n=\tau$. In this case, the coupling is constant
during each delay interval, and it is straightforward to integrate
Eq. \eqref{eq:linmodel}. We consider an arbitrary initial function
${\bf x}_{0}(t)$, $t\in[-\tau,0]$; decomposing into its Fourier
components

\begin{equation}
  {\bf x}_{0}(t)=\displaystyle\sum_{n=-\infty}^{+\infty}
  {\bf x}_{0n} e^{i\omega_n t}\,,
\end{equation}
with $\omega_n=2\pi n/\tau$, one finds for the evolution of the $n$-th
mode during the first delay interval

\begin{equation}
  \dot{\bf x}_{1n}(t) = -\lambda_0 {\bf x}_{1n} +
  \kappa A_1 {\bf x}_{0n}e^{i\omega_n t } \label{eq:linmodelfourier}\,,
\end{equation}
which is solved by

\begin{equation}
  {\bf x}_{1n}(t) = e^{-\lambda_0 t} {\bf x}_{0n} +
  \frac{\kappa}{\lambda_0+i\omega_n}
  \left(e^{i\omega_nt }-e^{-\lambda_0 t}\right)
  A_1{\bf x}_{0n}\label{eq:linsolfourier}\,.
\end{equation}
Since the terms proportional to $e^{-\lambda_0 t}$ become negligible
after a short transient of order $\mathcal{O}(\lambda_0^{-1})$, we can
approximate the general solution during the first delay interval as
${\bf x}_1(t)=\sum_n {\bf x}_{1n}e^{i\omega_n t}$, with

\begin{equation}
  {\bf x}_{1n} = \frac{\kappa}{\lambda_0+i\omega_n} A_1{\bf x}_{0n}
  \label{eq:linsolfourier2}\,.
\end{equation}
Note that, for a constant network, this corresponds to the decay rates
given by pseudocontinuous spectrum Eq. \eqref{eq:pseudospec}.

Repeating this procedure for $M$ time delays, and thus $M$
alternations of the topology, one finds a general solution
${\bf x}_M(t)=\sum_n {\bf x}_{Mn}e^{i\omega_n t}$, with

\begin{equation}
  {\bf x}_{Mn} =
  \left[\frac{\kappa}{\lambda_0+i\omega_n}\right]^M
  \left(\displaystyle\prod_{m=1}^M A_m\right){\bf x}_{0n}
  \label{eq:slownetwork}\,.
\end{equation}
Thus, we retrieve an ``effective'' adjacency matrix

\begin{equation}
A_\eff=\left(\displaystyle\prod_{m=1}^M A_m \right)^{1/M}\label{eq:slow}\,.
\end{equation}

\begin{figure}
  \includegraphics[width=8cm]{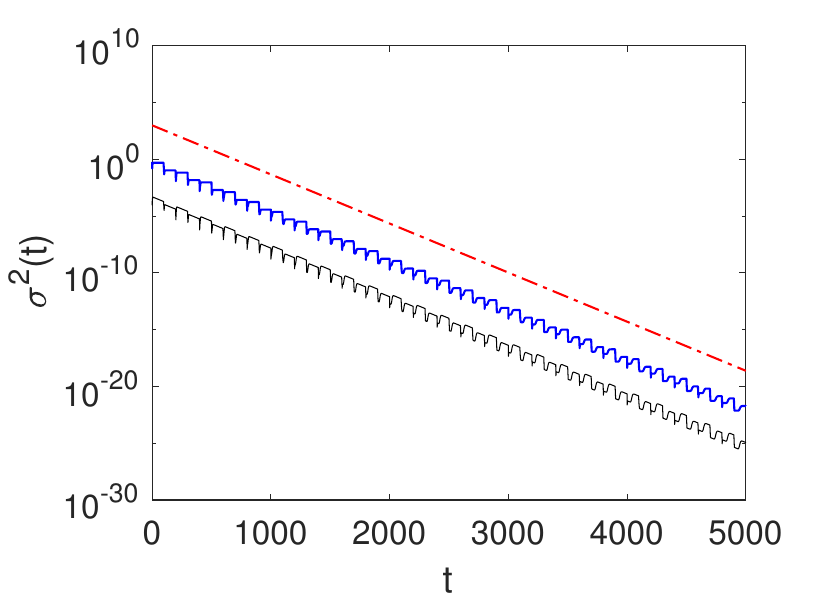}
   \caption{Evolution of the variance of the linear system Eq. \eqref{eq:linmodel} the systems topologies alternating between the matrices given in Eq. \eqref{eq:comm_choice}. For the lower full black line, initial conditions were exponentially decaying, ${\bf x}(t<0)={\bf x}_0 e^{-\lambda t}$, with $\lambda$ given by the theoretical prediction Eq. \eqref{eq:trans_lam}. For the middle full blue line, we used constant initial conditions. The black line has been shifted downwards for better comparison. The theoretical decay rate (shifted upwards) (Eq.  \eqref{eq:trans_lam}) is shown for comparison (upper red dotdashed line), the agreement is excellent. Parameters are $N=3$, $\kappa=0.8$, $\lambda_0=1$, $T_n=\tau=100$.}
  \label{fig:slownetworklimit}
\end{figure}

This prediction is verified numerically in Fig. \ref{fig:slownetworklimit}. Again, we simulated the model Eq. \eqref{eq:linmodel} for three nodes, with the coupling configuration alternating regularly between the commuting matrices given in Eq. \eqref{eq:comm_choice}. Our theory predicts a transverse decay rate given by Eq. \eqref{eq:trans_lam}), with, in the slow network limit, $\gamma_\eff$ given by the eigenvalues of the effective adjacency matrix Eq. \eqref{eq:slow}, 
$$\gamma_\eff=\sqrt{\gamma_1\gamma_2}\,,$$
with $\gamma_1$ and $\gamma_2$ the respective transverse eigenvalues of $A_1$ and $A_2$, as before. Fig. \ref{fig:slownetworklimit} compares the variance of the nodes for two different initial conditions (blue and black curves) with the theoretical prediction. Also in this case the theoretical prediction provides an excellent approximation for the numerical decay rates, the difference between the best linear fit of the simulations and the theoretical decay rate is around 1\%.

Note that the arithmetic mean network synchronizes faster than the geometric mean network. This is the case for most pairs of stochastic matrices. In the fast switching approximation typically the network synchronizes faster than both topologies between which it alternates, and thus the network fluctuations can be said to enhance synchronization. In contrast, one usually finds a transverse decay rate that is in between the decay rates of both topologies in the slow network limit. This is always the case if the adjacency matrices commute. Thus, an appropriate choice of topologies allows to control the synchronisation properties, also in the slow network limit.


\section{Interplay of time scales}
\label{sec:timescales}

The linear model given in Eq. \eqref{eq:linmodel} is difficult to solve in general. Nonetheless, it is possible to find exact solutions for the special case that the delay time $\tau$ is a multiple of the network time $T_n$. In the following, as a first simplification, we consider a regular alternation between two topologies, with respective adjacency matrices $A_1$ and $A_2$. Thus, the system repeats the cycle exactly every $2T_n$. The results can be generalised to a periodic sequence of topologies $A_1, \hdots, A_M$.

We assume a constant initial function ${\bf x}(t<0)={\bf x}_0$, since for
constant coupling, and in the slow network limit (see
Eqs. \eqref{eq:pseudospec} and \eqref{eq:slownetwork}) the zeroth
Fourier mode is the least stable, and therefore becomes crucial to
determine the stability. Similarly to the slow network limit, it is
possible to integrate Eq. \eqref{eq:linmodel} for consecutive delay
intervals, concatenating the output of an interval with the input for
the next. During the first delay interval, the dynamics is modelled as

\begin{eqnarray}
  \dot{\bf x}_1(t) & = &
  -\lambda_0 {\bf x}_1(t) +\kappa A_{1,2} {\bf x}_0\label{eq:genmodel}\,.
\end{eqnarray}
After an initial transient of the order of
$\mathcal{O}(\lambda_0^{-1})$, the resulting dynamics during the delay
interval is $2T_n$-periodic: the system evolves exponentially between
$t=kT_n$ and $t=(k+1)T_n$, when it changes direction. The values at
these turning points are denoted $\frac{\kappa}{\lambda_0}{\bf
  x}_{1A}$ and $\frac{\kappa}{\lambda_0}{\bf x}_{1B}$ respectively.

During the first half cycle of its periodic motion, the solution ${\bf
  x}_{A1}(t)$ of Eq. \eqref{eq:genmodel} then reads

\begin{eqnarray}
  {\bf x}_{A1}(t) & = & \frac{\kappa}{\lambda_0}
  \left[e^{-\lambda_0t} {\bf x}_{1A}+(1-e^{-\lambda_0t}) A_1{\bf x}_0
    \right]
  \label{eq:gensol1A}\,.
\end{eqnarray}

During the second half cycle of its periodic motion, the solution
${\bf x}_{B1}(t)$ of Eq. \eqref{eq:genmodel} reads

\begin{eqnarray}
  {\bf x}_{B1}(t) & = &
  \frac{\kappa}{\lambda_0}
  \left[e^{-\lambda_0t}{\bf x}_{1B}+(1-e^{-\lambda_0t})A_2{\bf x}_0
    \right]
  \label{eq:sol1B}\,.
\end{eqnarray}

Assuming a continuous periodic motion ${\bf
  x}_{B1}(T_n)=\frac{\kappa}{\lambda_0}{\bf x}_{1A}$ and ${\bf
  x}_{A1}(T_n)=\frac{\kappa}{\lambda_0}{\bf x}_{1B}$, we find a
solution for the turning points


\begin{equation}
  {\bf x}_{1A,B}  =  \frac{1}{2}\left((A_1+A_2)\mp
  \frac{1-e^{-\lambda_0T_n}}{1+e^{-\lambda_0T_n}}(A_1-A_2)\right){\bf x}_0\,.
\end{equation}
Note that in the limit $\lambda_0 T_n\ll 1$, we find as a leading
order approximation

\begin{equation}
  {\bf x}_{1A}={\bf x}_{1B}=\frac{1}{2}(A_1+A_2){\bf x}_0\,,
\end{equation}
which corresponds to the fast switching approximation. However, in
order to find the long term evolution, one needs to integrate over the
next delay intervals.

\subsection{Asymptotic behavior for $\tau=2M T_n$}

Clearly, the evolution depends on the interaction of the two coupling topologies. In the modelling equations, the driving terms in next delay intervals depend on combinations of both adjacency matrices. Thus, we expect a periodic (parity) effect in the overall transverse decay rate with respect to 
$\mod(\tau,2T_n$).

Let us focus on the case when $\tau$ is an even multiple of the
network time $T_n$, $\tau=2M T_n$, with $M\in\N$ arbitrarily
large. By solving Eq. \eqref{eq:linmodel} for the second delay interval
we obtain

\begin{eqnarray}
  \dot{\bf x}_{A,B2}(t) & = & -\lambda_0 {\bf x}_{A,B2}(t) +
  \kappa A_{1,2}{\bf x}_{A,B1}(t)\label{eq:2gen}\,.
\end{eqnarray}
Again, the system behaves periodically (after a few transients), now
between switching points $\frac{\kappa^2}{\lambda_0^2}{\bf x}_{2A}$
and $\frac{\kappa^2}{\lambda_0^2}{\bf x}_{2B}$. We find for the two
segments of the periodic motion

\begin{eqnarray}
  {\bf x}_{A,B2} (t) & = &
  \frac{\kappa^2}{\lambda_0^2}
  \left[\left(1-e^{-\lambda_0 t}(1+\lambda_0 t)\right)A_{1,2}^2 {\bf x}_0
    + \right.\nonumber\\
    & & \left.\lambda_0 t e^{-\lambda_0 t} A_{1,2}{\bf x}_{1A,B}
    + e^{-\lambda_0 t}{\bf x}_{2A,B}\right]\,.
\end{eqnarray}
Using continuity, we can solve for the turning points ${\bf x}_{2A}$
and ${\bf x}_{2B}$

\begin{widetext}
\begin{small}
\begin{equation}
  {\bf x}_{2A,B}  =
  \frac{1}{2}\left[\left(1-e^{-\lambda_0T_n}(1+\lambda_0 T_n)\right)
    \left(\frac{A_1^2+A_2^2}{1-e^{-\lambda_0 T_n}} \mp
    \frac{A_1^2-A_2^2}{1+e^{-\lambda_0 T_n}}\right) {\bf x}_0 +
    e^{-\lambda_0 T_n} \lambda_0 T_n
    \left(\frac{A_1 {\bf x}_{1A}+A_2 {\bf x}_{1B}}{1-e^{-\lambda_0 T_n}}
    \mp \frac{A_1 {\bf x}_{1A}-A_2 {\bf x}_{1B}}{1+e^{-\lambda_0 T}}\right)\right]\,.
\end{equation}
\end{small}
\end{widetext}

Repeating this procedure $n$ times, we find the following recursive
relation for the turning points $\frac{\kappa^n}{\lambda_0^n}{\bf
  x}_{nA,B}$ in the $n$-th delay interval,

\begin{widetext}
\begin{eqnarray}
  {\bf x}_{nA,B} &  =  &
  \frac{1}{2}\left[\left(1-e^{-\lambda_0T_n}\sum_{k=0}^{n-1}
    \frac{(\lambda_0 T_n)^k}{k!}\right)
    \left(\frac{A_1^n+A_2^n}{1-e^{-\lambda_0 T_n}}
    \mp \frac{A_1^n-A_2^n}{1+e^{-\lambda_0 T_n}}\right) {\bf x}_0 \right.
    \nonumber\\
    & &  \left. + e^{-\lambda_0 T_n}
    \sum_{k=1}^{n-1}\frac{(\lambda_0 T_n)^k}{k!}
    \left(\frac{A_1^k {\bf x}_{(n-k)A}
      +A_2^k {\bf x}_{(n-k)B}}{1-e^{-\lambda_0 T_n}} \mp
  \frac{A_1^k {\bf x}_{(n-k)A}-A_2^k {\bf x}_{(n-k)B}}{1+e^{-\lambda_0 T}}
  \right)\right]
  \label{eq:ngen}\,.
\end{eqnarray}
\end{widetext}

In the fast switching approximation $\lambda_0T_n \rightarrow 0$ one
retrieves straightforward the average network solution

\begin{equation}
{\bf x}_{nA}={\bf x}_{nB}\equiv {\bf
  x}_n=\frac{A_1+A_2}{2}{\bf x}_{n-1}
\end{equation}
as the leading order term (for any value of $n$) .

If the network time is larger or of the same order as the internal
time scale, $\lambda_0 T > 1$, the first term of Eq. \eqref{eq:ngen}, which is proportional to
$\left(1-e^{-\lambda_0T_n}\sum_{k=0}^{n-1} \frac{(\lambda_0
  T_n)^k}{k!}\right)$ is dominant during the first few delay
intervals. For slow networks $\lambda_0 T_n \gg 1$, this results in an
initial decay rate depending on the largest transverse eigenvalue of
any of the two matrices $A_1$ or $A_2$. However, as time and $n$
increase, the first term tends to zero: it becomes negligible at
$n\approx 4 \lambda T_n$.


In this limit $n> 4\lambda_0 T_n$, we find an approximate solution for
Eq. \eqref{eq:ngen},

\begin{equation}
  {\bf x}_{nA,nB} \approx A^n {\bf x}_{0A,0B}\,.
\end{equation}
with the effective adjacency matrix $A$ given by
\begin{eqnarray}
A & = & \frac{1}{2}(A_1+A_2)\label{eq:Tneven}\,,
\end{eqnarray}
upon the condition that $A$ is invertible. Thus, we find that the arithmetic mean network determines the asymptotic decay rate, independent of $T_n$, and even though the fast network limit does not apply. However, the synchronization time, and -in a nonlinear system- the basin of attraction, depends as well on the initial evolution rate, which depends in turn on the network time $T_n$.

\begin{figure}
  \includegraphics[width=8cm]{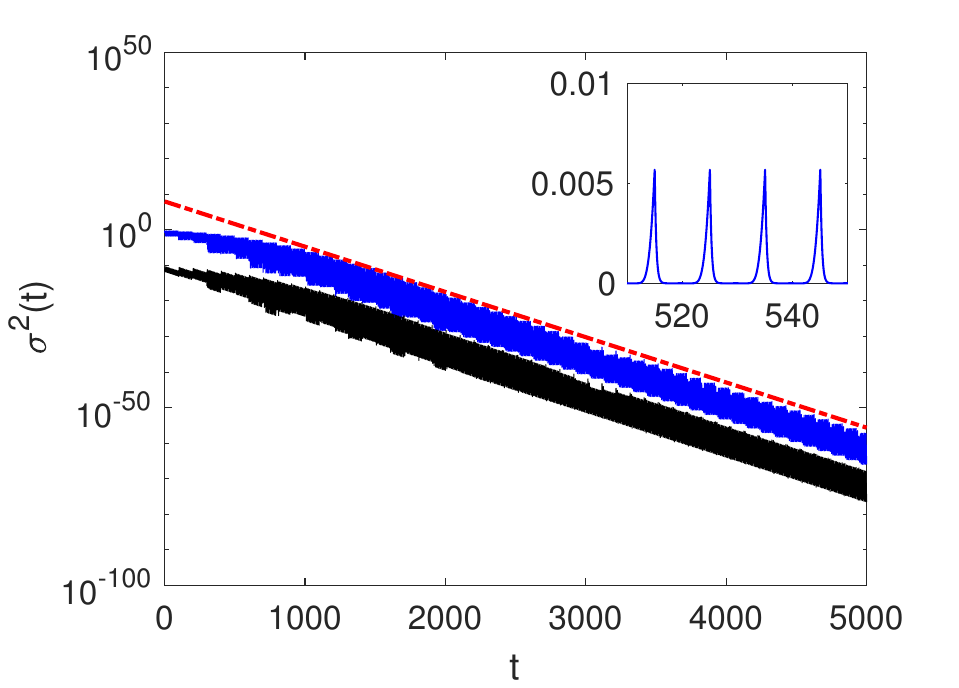}
   \caption{Evolution of the variance of the linear system Eq. \eqref{eq:linmodel} with the systems topologies alternating between the matrices given in Eq. \eqref{eq:comm_choice}. For the lower full black line, initial conditions were exponentially decaying, ${\bf x}(t<0)={\bf x}_0 e^{-\lambda t}$, with $\lambda$ given by the theoretical prediction Eq. \eqref{eq:trans_lam} for the fast network limit. For the middle full blue curve we used constant initial conditions, in agreement with our assumption for the analytical calculations. The blue line has been shifted downwards for better comparison. The theoretical decay rate (shifted upwards) (Eqs.  \eqref{eq:trans_lam}, Eq. \eqref{eq:Tneven}) is shown for comparison (upper red dotdashed line). The inset shows the $2T_n$-periodic oscillations of the variance during the fifth delay interval. Parameters are $N=3$, $\kappa=0.8$, $\lambda_0=1$, $T_n=5$, $\tau=100$.}
  \label{fig:Tneven}
\end{figure}

We have simulated the system in Fig. \ref{fig:Tneven}. We again consider three nodes modelled by Eq. \eqref{eq:linmodel}, which alternate regularly between the commuting matrices given in Eq. \eqref{eq:comm_choice}. We chose $\tau=20 T_n$. For constant initial conditions (blue curve), we show in the inset the $2 T_n$-periodic behavior within each delay interval that we model analytically, the main panel illustrates the decay of the variance.  For our choice of $T_n=5$, one can distinguish between the slower initial decay and the long term faster decay rate. Also for exponential initial conditions (with the exponent given by the fast network approach), plotted in black, we observe similar trends, with initially slow decay, and fast decay on the long term, in correspondence with Eq. \eqref{eq:ngen}. 

The theoretical transverse decay rate is given by Eq. \eqref{eq:trans_lam}), with $\gamma_\eff$ in the long time given being the maximal transverse eigenvalue of the effective adjacency matrix (Eq. \eqref{eq:Tneven}). Although the variance fluctuates stronger in this case than in the slow and fast network limits, the agreement between theoretical approximation and numerical simulations is good. 

\subsection{Asymptotic decay rate for $\tau=(2M+1) T_n$}

If the delay time is an odd multiple of the network time, the
  topologies interfere in a different way: the adjacency matrix $A_1$ enters
  the driving term multiplying the coupling matrix $A_2$. In this
case, the modelling equations during the second delay interval read

\begin{eqnarray}
  \dot{\bf x}_{A,B2}(t) & = &
  -\lambda_0 {\bf x}_{A,B2}(t) + \kappa A_{1,2}{\bf x}_{B,A1}(t)\,.
\end{eqnarray}
Note that, in the delayed input term ${\bf x}_{A1}(t)$ and ${\bf
  x}_{B1}(t)$ have switched role, compared to the even multiple case
(Eq. \eqref{eq:2gen}). This results in turning points ${\bf x}_{2A}$
and ${\bf x}_{2B}$, given by

\begin{widetext}
\begin{small}
\begin{equation}
  {\bf x}_{2A,B}  =
  \frac{1}{2}\left[\left(1-e^{-\lambda_0T_n}(1+\lambda_0 T_n)\right)
    \left(\frac{A_1A_2+A_2A_1}{1-e^{-\lambda_0 T_n}}
    \mp \frac{A_1A_2-A_2 A_1}{1+e^{-\lambda_0 T_n}}\right) {\bf x}_0
    + e^{-\lambda_0 T_n} \lambda_0 T_n
    \left(\frac{A_2 {\bf x}_{1A}+A_1 {\bf x}_{1B}}{1-e^{-\lambda_0 T_n}}
 \mp \frac{A_2 {\bf x}_{1A}-A_1 {\bf x}_{1B}}{1+e^{-\lambda_0 T}}\right)\right]\,.
\end{equation}
\end{small}
\end{widetext}

In the $n$-th delay interval ($n$ even), the the turning points
$\frac{\kappa^n}{\lambda_0^n}{\bf x}_{nA,nB}$ are given by

\begin{widetext}
\begin{eqnarray}
  {\bf x}_{nA,B} & = &
  \frac{1}{2}\left(1-e^{-\lambda_0T_n}\sum_{k=0}^{n-1}
  \frac{(\lambda_0 T_n)^k}{k!}\right)
  \left(\frac{(A_1A_2)^{n/2}+(A_2A_1)^{n/2}}{1-e^{-\lambda_0 T_n}}
  \pm \frac{(A_1A_2)^{n/2}-(A_2A_1)^{n/2}}{1+e^{-\lambda_0 T_n}}\right) {\bf x}_0
  \nonumber\\
  & & + \frac{1}{2} e^{-\lambda_0 T_n}
  \left[\sum_{k=0}^{(n-2)/2}\frac{(\lambda_0 T_n)^{2k+1}}{(2k+1)!}\times\right.
    \label{eq:n2gen}\\
& & \left(\frac{A_2 (A_1A_2)^k {\bf x}_{(n-2k-1)A}+A_1(A_2A_1)^k {\bf x}_{(n-2k-1)B}}{1-e^{-\lambda_0 T}}\pm \frac{A_2(A_1A_2)^k {\bf x}_{(n-2k-1)A}-A_1(A_2A_1)^k {\bf x}_{(n-2k-1)B}}{1+e^{-\lambda_0 T}}\right)\nonumber\\
& & + \left.\sum_{k=1}^{(n-2)/2}\frac{(\lambda_0 T_n)^{2k}}{(2k)!}\left(\frac{(A_1A_2)^k{\bf x}_{(n-2k)A}+(A_2A_1)^k{\bf x}_{(n-2k)B}}{1-e^{-\lambda_0 T}}\pm \frac{(A_2A_1)^k{\bf x}_{(n-2k)B}-(A_1A_2)^k{\bf x}_{(n-2k)A}}{1+e^{-\lambda_0 T}}\right)\right]\nonumber\,.
\end{eqnarray}
\end{widetext}
If $\lambda_0 T_n$ is small, we retrieve the fast switching approximation. Just like in the even-numbered case, for large $\lambda_0 T$ the first
term part remains dominant during the first few delay intervals. This
leads to a decay rate that relates to the spectral gap of $A_1 A_2$;
if $A_1$ and $A_2$ commute, the initial decay rate depends on the
eigenvalues of the product $A_1A_2$.
In the long time limit, it is possible to find an asymptotic solution
Eq. \eqref{eq:n2gen} for large $\lambda_0 T_n$, under the condition
that both adjacency matrices commute. Along a transverse eigenvector
$x$, the respective eigenvalues of $A_1$ and $A_2$ are then denoted
$\gamma_1$ and $\gamma_2$. In this case, Eq. \eqref{eq:n2gen}
simplifies to

\begin{eqnarray}
x_{nA,B} & = & 
e^{-\lambda_0 T_n}
\left[\sum_{k=0}^{(n-2)/2}\frac{(\lambda_0 T_n)^{2k+1}}{(2k+1)!}
  \gamma_{2,1} \tilde{\gamma}^{2k} x_{(n-2k-1)A,B}\right.
  \nonumber\\ 
  &  &\left. + \sum_{k=1}^{(n-2)/2}
  \frac{(\lambda_0 T_n)^{2k}}{(2k)!}
  \tilde{\gamma}^{2k} x_{(n-2k)B,A}\right] \,,
\end{eqnarray}
with $\tilde{\gamma}=\sqrt{\gamma_1\gamma_2}$, and choosing the sign such that the difference between the phases $\arg(\tilde{\gamma}$ and $\arg(\gamma_1+\gamma_2)$ is minimal. This allows a solution

\begin{equation}
x_{nA,B}=\gamma^n x_{0A,0B}\,,
\end{equation}
with the effective eigenvalue $\gamma$ given by

\begin{equation}
  \gamma =\frac{\tilde{\gamma}}{1+\frac{1}{\lambda_0 T_n}
    (\ln[\tilde{\gamma}]-\ln[(\gamma_1+\gamma_2)/2])}\label{eq:Tnodd}\,.
\end{equation}

For $\tau$ being an odd multiple of $T_n$, the decay rate shows an evolution from the fast network limit for small $T_n$ to the slow network limit for large $T_n$, with a linear dependence on $(\lambda_0 T_n)^{-1}$ in the latter case. In the particular case that $|\kappa\tilde{\gamma}|=\lambda_0$, we retrieve a power law behavior $\lambda\propto T_n^{-1}$, as is demonstrated in Section \ref{sec:simlin}.

\begin{figure}
  \includegraphics[width=8cm]{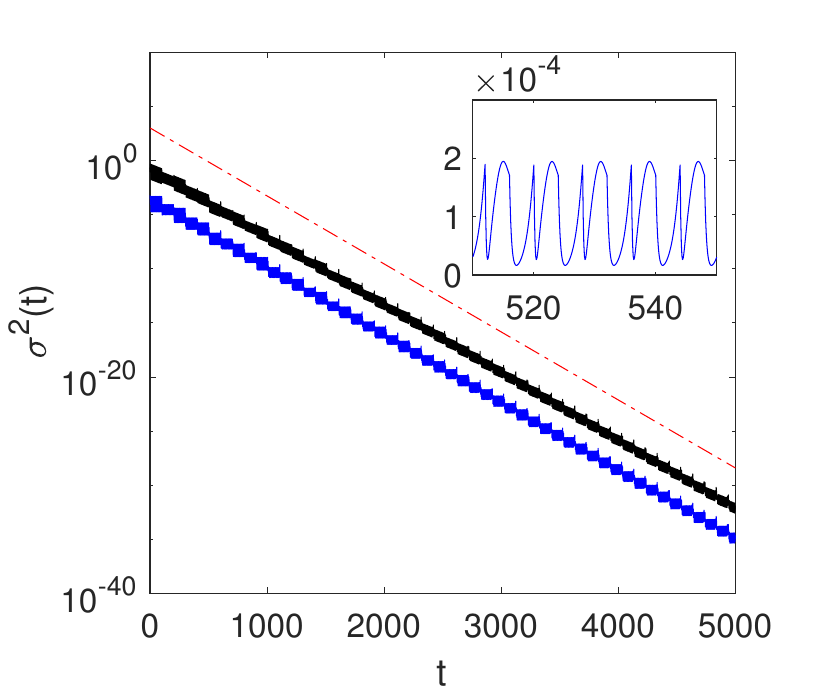}
   \caption{Evolution of the variance of the linear system Eq. \eqref{eq:linmodel} with the systems topologies alternating between the matrices given in Eq. \eqref{eq:comm_choice}. For the middle full black line, initial conditions were exponentially decaying, ${\bf x}(t<0)={\bf x}_0 e^{-\lambda t}$, with $\lambda$ given by the theoretical prediction Eq. \eqref{eq:trans_lam} for the slow network limit. For the lower full blue line, we used constant initial conditions, in agreement with our assumption for the analytical calculations. The blue line has been shifted downwards for better comparison. The theoretical decay rate (shifted upwards) (Eqs.  \eqref{eq:trans_lam}, \eqref{eq:Tnodd}) is shown for comparison (upper red dotdashed line). The inset shows the $2T_n$-periodic oscillations of the variance during the fifth delay interval. Parameters are $N=3$, $\kappa=0.8$, $\lambda_0=1$, $T_n=4$, $\tau=100$.}
  \label{fig:Tnodd}
\end{figure}

We compare these theoretical results to numerical simulations in Fig. \ref{fig:Tnodd}. The network Eq. \eqref{eq:linmodel} of three nodes, alternates regularly between the commuting matrices given in Eq. \eqref{eq:comm_choice}, with $\tau=25 T_n$. For constant initial conditions (blue curve), the $2 T_n$-periodic behavior within each delay interval is exemplified in the inset, while the main panel shows the overall exponential decay.  Moreover, we find that this decay rate does not change significantly for non-constant initial functions (black curve), hence demonstrating the validity of our analytic approach.  

The theoretical transverse decay rate is given by Eq. \eqref{eq:trans_lam}), with $\gamma_\eff$ given by Eq. \eqref{eq:Tnodd}, with $\gamma_1$ and 
$\gamma_2$ the respective transverse eigenvalues of $A_1$ and $A_2$, and $\tilde{\gamma}=\sqrt{\gamma_1\gamma_2}$. We find again excellent agreement between theory and simulations. Note the considerably different transverse decay rates in Figs. \ref{fig:Tneven} and \ref{fig:Tnodd}, while the network times do not differ much ($T_n=5$ and $T_n=4$, respectively), illustrating the sensitive dependence of the system on the ratio of time scales.

\section{Simulations for varying network time}
\label{sec:simlin}

To explore the synchronization properties in the full range of network times $T_n$, we have performed numerical simulations of the aforementioned linear system Eq. \eqref{eq:linmodel} with delayed interactions:

$$
\dot x(t)=-\lambda_0 x(t) + \kappa A(t) x(t-\tau)\,.
$$
%
In order to easily compare with the analytic results, we consider $A(t)$ to be a discontinuous matricial process which proceeds through the
alternation of two matrices, $A_1$ and $A_2$, every $T_n$. The initial function in our simulations
is provided by fixing a certain $x(t)=x_0$ for all $t<0$.

\begin{figure}
  \includegraphics[width=8cm]{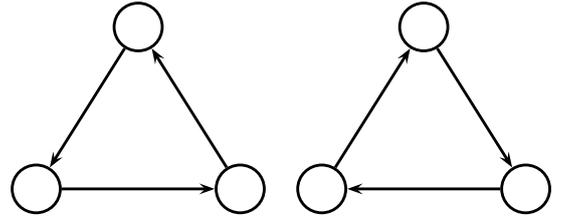}
  \caption{Illustration of the two topologies giving rise to the
    cyclic matrices, $A_1$ and $A_2$.}
  \label{fig:illust_cyclic}
\end{figure}

Unless otherwise mentioned, our network consists of three nodes, $N=3$. Our first choice for $A_1$ and $A_2$ is the {\em cyclic} choice,
which is illustrated in Fig. \ref{fig:illust_cyclic}. The system alternates between
the two different cyclic directed graphs between three nodes, i.e.:

\beq
A_1=\begin{pmatrix} 0 & 1 & 0 \\ 0 & 0 & 1 \\ 1 & 0 & 0
\end{pmatrix},\quad
A_2=\begin{pmatrix} 0 & 0 & 1 \\ 1 & 0 & 0 \\ 0 & 1 & 0
\end{pmatrix}.
\label{eq:cyclic_choice}
\eeq

Note that these matrices commute and, moreover, they are
inverse, i.e. $A_1A_2=A_2A_1=I$.

In a second set of simulations, the {\em commuting} choice, we use the topologies $A_1$ and $A_2$ already introduced in Eq. \eqref{eq:comm_choice}. In this case all theoretical results can be applied, but the matrices are no longer inverse. A third choice for the topologies $A_1$ and $A_2$ is the {\em random} choice, where we select always $A_{ij}$ as independent uniform deviates in $[0,1]$, and impose the unit row-sum condition afterwards. In this case the two adjacency matrices do not commute. 

\begin{figure}
  \includegraphics[width=8cm]{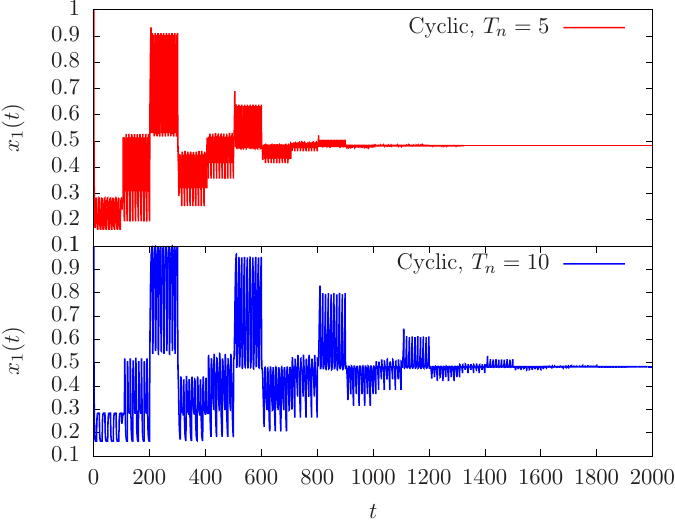}
  \includegraphics[width=8cm]{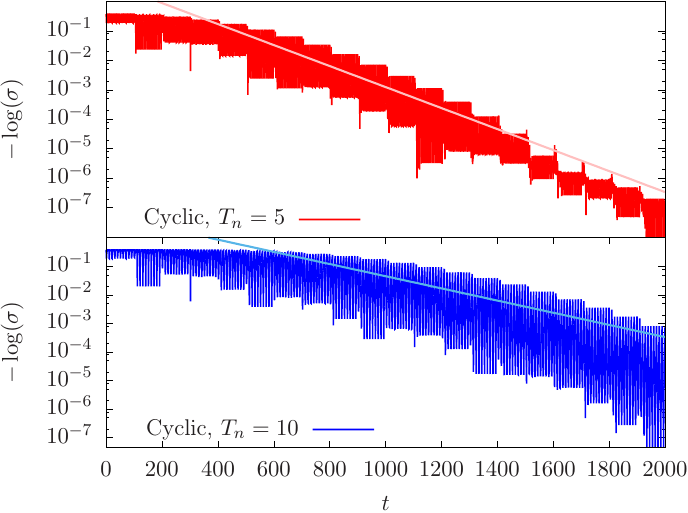}
  \caption{Two histories of Eq. \eqref{eq:linmodel} for the cyclic
    choice of $A_1$ and $A_2$, given by Eq. \eqref{eq:cyclic_choice}. Top:
    Evolution of a single component. Bottom: evolution of the minus
    the logarithm of the deviation. Always using $N=3$,
    $\lambda_0=\kappa=1$, $\tau=100$, $T_n=5$ and $T_n=10$. The straight
    lines correspond to the best linear fit, allowing us to estimate
    the transversal decay rate (TDR).}
  \label{fig:history}
\end{figure}

Fig. \ref{fig:history} shows a few illustrative histories when 
two cyclic matrices are alternated along Eq. \eqref{eq:linmodel}, using
$\lambda_0=\kappa=1$, $\tau=100$ and two values for the network
switching time: $T_n=10$ and $T_n=5$, and a fixed random initial
condition. The delay equations are integrated using $\Delta
t=10^{-2}$. The top panel shows the evolution of the first component,
$x_1(t)$. In both cases, the delay time $\tau$ is an even multiple of the network time $T_n$. The similarities with Fig. \ref{fig:Tneven}, which shows an alternation of different topologies, for different parameters, but for the same ratio of time scales, are clear.

As predicted by Eq. \eqref{eq:Tneven} for our choice of parameters, in all the cases shown the different components 
approach a synchronized state. 

The second panel of
Fig. \ref{fig:history} shows the evolution of minus the logarithm of
the deviation between all components, along with our estimate for an
exponential decay. The slower decay for $T_n=10$ results from the larger transient time. Notice that the precise measurement of the decay rate is
hindered by the strong oscillations, whose periodicity is given by the
interaction delay $\tau$. In order to measure it properly in a robust
way, we select a random sample of points along the evolution. We do
not use a fixed interval sample in order to avoid lattice
artifacts. Then, we select random sub-samples of those data, fitting
each of them to a straight line. The average of the slopes is our
estimate for the TDR, while the standard deviation of these values
provides an estimate of our uncertainty.

Using this method, we have computed the transverse decay rate in a variety of cases.  Our theoretical estimates for the TDR are
provided by Eq. \eqref{eq:trans_lam}, where the value of $\gamma_\eff$
corresponds to the arithmetic mean of the matrices in the
fast-switching regime and to the geometric mean in the slow-switching,
which corresponds to $T_n \sim \tau$. 

\begin{figure}
  \includegraphics[width=8cm]{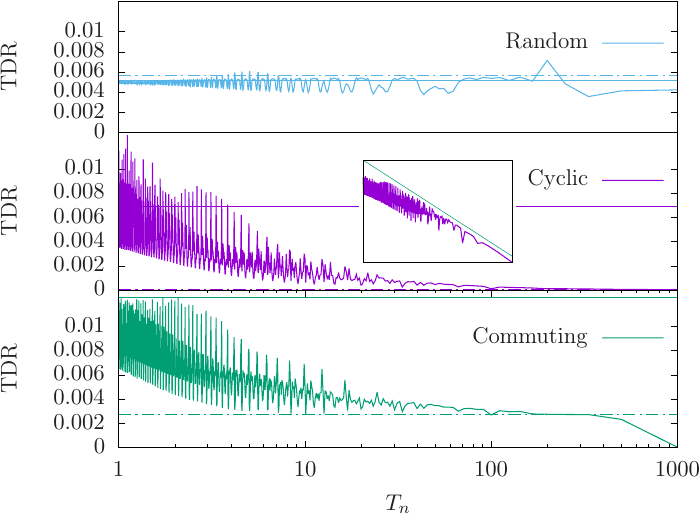}
  \includegraphics[width=8cm]{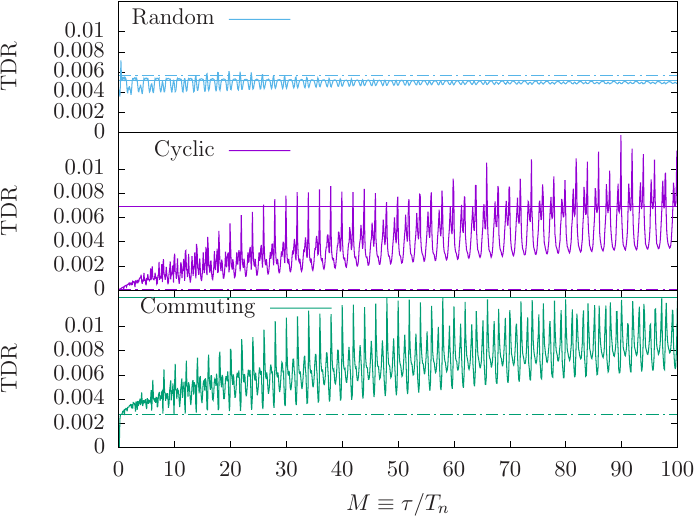}
  \caption{TDR of linear systems following Eq. \eqref{eq:linmodel} with
    $N=3$, $\tau=100$, $\lambda_0=\kappa=1$ for three different
    systems topologies: (a) random; (b) cyclic choice, given by
    Eq. \eqref{eq:fixed_choice}; (c) commuting choice, given by
    Eq. \eqref{eq:comm_choice}. The horizontal lines mark our
    theoretical estimates in the fast (dashed) and slow (continous)
    regimes. Top panel: TDR as a function of $T_n$. Notice the mild
    variations for (b) and (c), compared to the power-law behavior of
    (a): TDR $\sim T_n^{-1}$. Bottom panel: TDR as a function of
    $M\equiv \tau/T_n$. Notice the strong parity oscillations. }
  \label{fig:small}
\end{figure}

The top panel of Fig. \ref{fig:small} shows our estimates for the TDR
as a function of $T_n$ using $\tau=100$,
$\lambda_0=\kappa=1$. Each panel is devoted to a different choice for
the $A_1$ and $A_2$ matrices: random (top), cyclic (central) and
commuting (bottom). The horizontal lines mark our theoretical estimate
in the fast-switching (continuous line) and slow-switching (dashed
line) regimes. For the random case chosen (top panel), where matrices
$A_1$ and $A_2$ do {\em not} commute, we observe a good correspondence to the fast switching limit for small values of $T_n$, and a correspondence to the slow network case for a range of network times $T_n\approx \tau$, but there is no general trend visible. This may be due to the proximity of both limits.

In the commuting cases (central and bottom), however, there is a clear trend visible. Again, for fast fluctuations the transverse stability is well approximated by the fast switching approximation. The TDR evolves in both cases towards the slow network limit as $T_n\sim\tau$, the general trend is well described by decay rate for odd multiples (Eq. \eqref{eq:Tnodd}). For the commuting and random choices, the estimate for the decay rate decreases further as $T_n\gg\tau$. This can be explained by the fact that the dynamics, operates on a time scale $\tau$ and thus adiabatically follows the temporary network, which evolves on a slower time scale in this limit.

For our choice of parameters, the slow-switching regime in the cyclic case provides a TDR $\sim 0$, as shown in our data. In this case the
decay from the fast to the slow switching regimes responds approximately to a power-law, with TRD $\lambda \sim T_n^{-1}$, as shown in
the inset of Fig. \ref{fig:small} (middle). This power law can be explained by the odd multiple case. Indeed, combining Eqs. \eqref{eq:Tnodd} and Eq. \eqref{eq:trans_lam}, we can deduce,
\begin{eqnarray}
\lambda & \approx & \frac{1}{\tau}\ln\left|\frac{\kappa\gamma_\eff}{\lambda_0}\right|\nonumber\\
 & \approx & \frac{1}{\tau}\left(\ln\left|\frac{\kappa\tilde{\gamma}}{\lambda_0}\right|-\ln\left[1-\frac{1}{\lambda_0 T_n}\ln[-1/2(\gamma_1+\gamma_2)]\right]\right)\nonumber\\
 & \approx & \frac{-1}{\tau}\frac{\ln 2}{\lambda_0 T_n}\,,
\end{eqnarray}
where we used $\tilde{\gamma}=-1$, our parameter choice $\kappa=\lambda_0=1$ and $(\lambda_0 T_n)^{-1}$ is considered small.

The bottom panel shows the same data, but with a restructured
abscissa: the new independent variable is $\tau/T_n\equiv M$. In this
case we the parity oscillations in the fixed and the cyclic
case are clearly visible. In particular, we observe synchronizing resonances corresponding to the even values of $M$, in agreement to the theoretical predictions. In the commuting case the TDR at these resonances is particularly well described by the arithmetic mean, also in agreement with theory (Eq. \eqref{eq:Tneven}.

\begin{figure}
  \includegraphics[width=8cm]{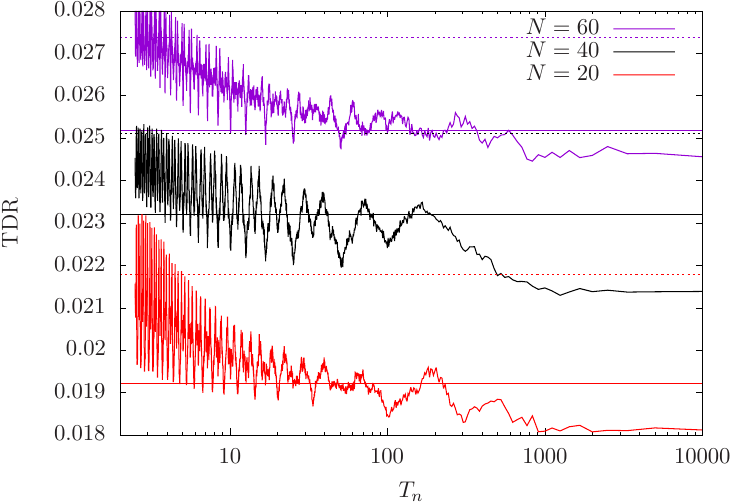}
  \includegraphics[width=8cm]{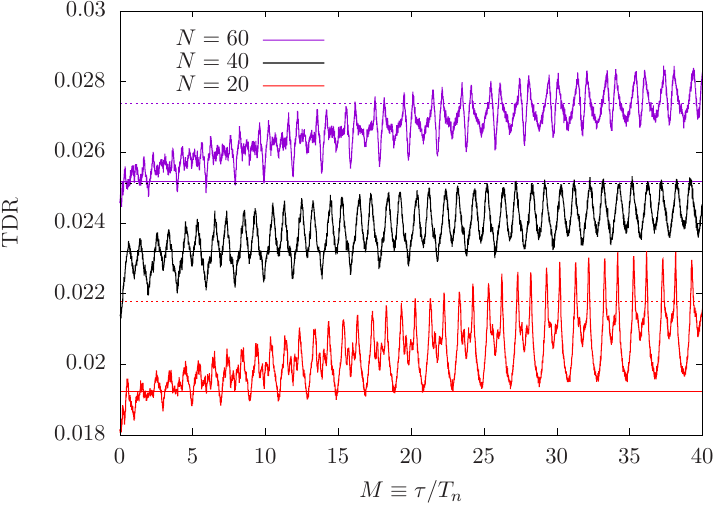}
  \caption{TDR as a function of $M=\tau/T_n$ for random choices of
    $A_1$ and $A_2$, and two different initial conditions. Notice the
    strong parity effect in all cases. For each case we have marked
    with an horizontal dashed line the theoretical prediction for the
    TDR in the large $M$ approximation (arithmetic mean, fast
    switching) and with a continuous line the small $M$ approximation
    (geometric mean, slow switching).}
  \label{fig:large}
\end{figure}

In Fig. \ref{fig:large} we check whether this evolution from arithmetic mean effective topology in the fast network limit to geometric mean in the slow network limit hold for
larger lattices with random (non-commuting) adjacency matrices, $A_1$
and $A_2$. We simulated networks of sizes $N=20$, $N=40$ and $N=60$ (see Fig. 8 from bottom to top, respectively). Again, the fast
and slow regime approximations to the TDR are marked with dashed and
continuous horizontal lines, respectively. We observe a good general
agreement between the theoretical prediction and the numerical data,
even though the matrices are non-commuting in this case. Moreover, we
observe the same strong parity oscillations.

\section{synchronization of delay-coupled chaotic maps}
\label{sec:simnonlin}

We illustrate our results for linear, time-continuous systems in a network of chaotic maps. Similar as in previous work \cite{Javi}, we consider a time-varying network of $N$ delay-coupled Bernouilli maps with a discrete time evolution modelled by

\begin{equation}
  {\bf x}(t+1) = (1-\epsilon)f({\bf x}(t)) + \epsilon A(t)f({\bf x}(t-\tau))\,,
  \label{eq:bernmodel}
\end{equation}
where ${\bf x}=(x_1, \dots, x_N)$ is the vector of the Bernoulli unit
states, $f(x)=a x \,\mod 1$, with $a>1$.  The network adjacency matrix $A(t)$
belongs to a sequence $\{A_1, A_2, \cdots\}$ of adjacency matrices
randomly sampled from a Newmann-Watts small-world network ensemble \cite{Newman-Watts}. Instances of this ensemble are directed
small-world networks similar to the standard Watts-Strogatz networks,
with a directed outside ring and each node has a probablity $p$ of
establishing a new ``shortcut'' link with a randomly chosen node.  The
difference with standard small-world networks is that here shortcuts are added without removing the
corresponding ring links, thus keeping the outside ring fixed and
ensuring the connectivity of the ensemble networks.

Like in the preceeding sections, the connectivity switches instantly every $T_n$ time-steps. The system's evolution to our linear model can be compared to the time-continous model Eq. \eqref{eq:linmodel}, by identifying the (opposite) sub-Lyapunov exponent and the internal decay rate. The coupling strength can simply be mapped,
\begin{equation}
  \lambda_0 = - \ln\left|1-a(1-\epsilon)\right|, \qquad \kappa=\epsilon a\,.
  \label{eq:paramsberntolinear}
\end{equation}

Notice that for this system, the instantaneous decay rate $\lambda_0$ is
dependent on the coupling strength $\epsilon$. For chaotic maps, the
transverse decay rate from the linear system coincides with the
so-called synchronization, or transverse Lyapunov exponent (SLE). That is, the rate
governing the evolution of small perturbations around the synchronized
state, $x_i(t)=x(t),\; \forall i$. In a fixed network of Bernoulli units, the SLE is, similar the linear system (Eq. \eqref{eq:pseudospec}), computed as follows \cite{Kanter} :
\begin{equation}
  \lambda=\frac{1}{\tau}\ln
  \left|\frac{a\epsilon\gamma_2}{1-a(1-\epsilon)}\right|\,,
  \label{eq:theobernlyap}
\end{equation}
where $\gamma_2$ is the adjacency matrix' second largest eigenvalue. 

\begin{figure}
  \includegraphics[width=\linewidth]{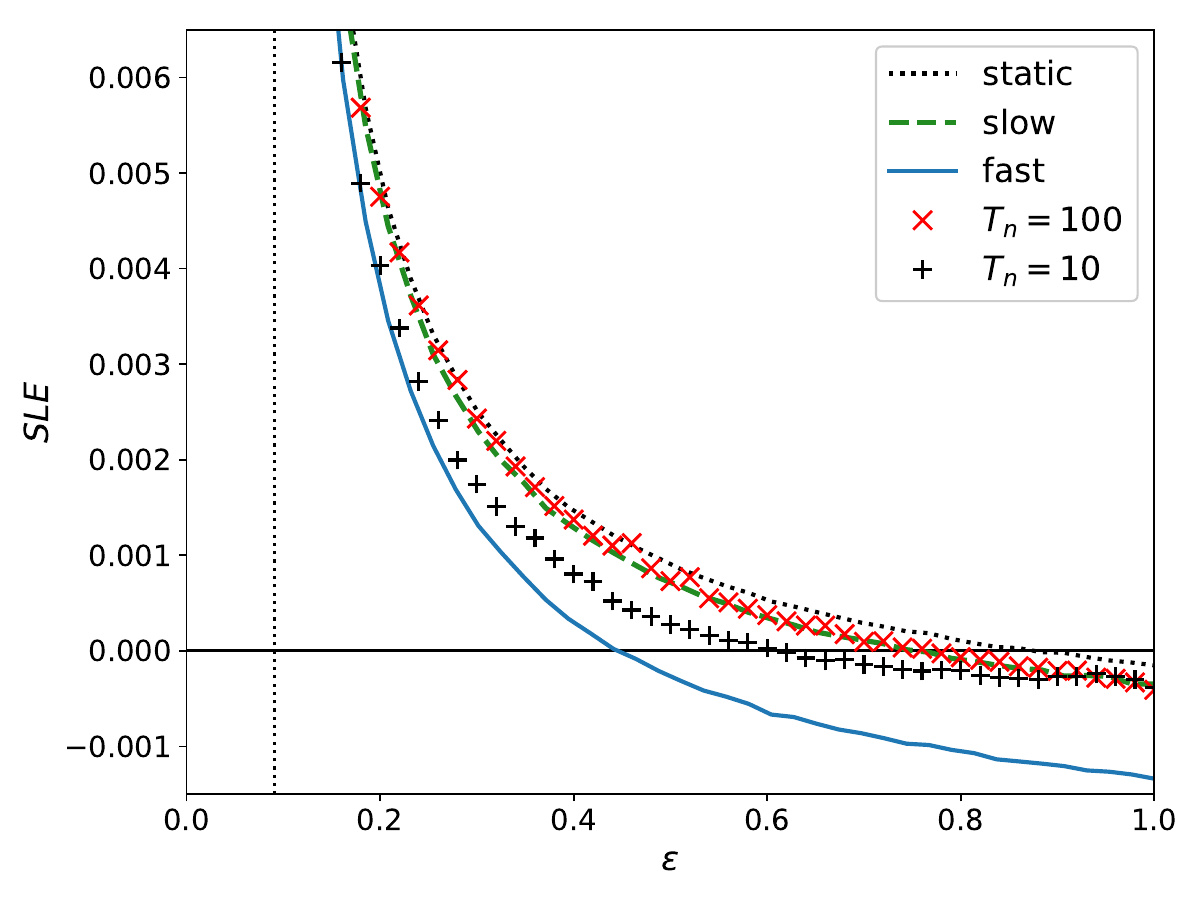}
  \caption{Numerical SLE corresponding to the evolution of a
    time-varying delay-coupled Bernouilli network with $N=40$, $\tau=100$,
    $a=1.1$, $p=0.5$. Each point is the average over $20$ simulated
    histories of $10^7$ time-steps. The solid blue and dashed green
    lines correspond to average SLE of the fast and slow effective
    networks, respectively, obtained as the arithmetic and geometric ensemble mean
    matrices. We also plot the average SLE of the
    static case, which is based on the mean transverse eigenvalue $|\gamma_2|$ of the ensemble. Both the slow
    and fast effective SLE enlarge the synchronization region
    ($SLE<0$) with respect to the static case. The vertical dashed
    line marks the limit of the weak chaos region, $\epsilon>(a-1)/a$.
  }
  \label{fig:bernSLE}
\end{figure}

We simulated the dynamics, Eq. \eqref{eq:bernmodel}, for several values
of the coupling strength $\epsilon$, with a fixed $p$ and for two
network switching times: $T_n=100$ and $T_n=10$. The comparison
between the resulting numerical SLE and the effective SLE
corresponding to the fast and slow network approximations is plotted
on figure \ref{fig:bernSLE}. The first case, $T_n=100$, corresponds to the
slow regime conditions $\lambda_0^{-1}\ll T_n=\tau$, and the numerical
SLE values follow the slow network approximation closely.  The second
case, $T_n=10$, crosses over from the fast network regime
$\lambda_0^{-1}\gg T_n$ at low $\epsilon$ values to the slow network
regime at values of $\epsilon\sim 1$. This is visible in the measured
SLE values, which evolves from the fast effective SLE to the slow
effective SLE as $\epsilon$ increases.


\section{Conclusion}
\label{sec:conclusions}

We have developed an analytical formalism within the linearized limit in order to understand synchronization phenomena in the case of delay-coupled networks with a fluctuating topology, studied previously by us in \cite{Javi}.

Considering a regular alternation of the topology between different configurations, we have studied both the case of fast and slow fluctuations. Based on the Master Stability Function approach, we derive the stability of a symmetric state based on the alternation of topologies and the interplay between the timescales involved: the internal timescale of the network nodes, the coupling delay, and the characteristic fluctuation time of the network. In two different regimes we have derived an ``effective''€ coupling topology: When the network fluctuations are faster than the internal time scale, the fast switching approximation can be extended to time-delay systems, and we find that the effective network is given by the arithmetic average. When the coupling delay and network fluctuation time are similarly large compared to the internal time scale of the network nodes , the effective network is given by the geometric average over the different topologies.

As the network time varies between the fast and slow limit, we find analytically a parity effect: the synchronization properties depend on the ratio of the delay time and the network time. In particular, if the adjacency matrices of the different topologies commute, one retrieves the fast network limit if the ratio between both time scales is even, while for odd ratios the behavior evolves from the fast to the slow network limit as the network time increases. Complementing these results with numerical simulations, we (broadly) recover this evolution from fast to slow network limit, from arithmetic to geometric mean network with increasing network time. This trend is visible for all network sizes, and for non-commuting and commuting topologies as well, but the agreement with the analytical theory is better in the latter case, and for larger networks. The parity effect is shown to be a universal feature for a regularly alternating topology. 

Finally, we compare our theoretical results to the synchronization properties of an ensemble of delay-coupled chaotic maps. As the coupling strength, and thus the internal time scale varies, we show that the synchronization boundaries shift between the geometric and arithmetic means of the ensemble, confirming the linear theory.

Future extensions of the research might be the study of other network
ensembles, such as random Erd\"os-R\'enyi graphs, scale-free networks
or even more complicated graphs of multiplex type with further
application to real world problems concerning transport and energy issues
or problems of supply networks in the general context of ``Smart
cities''.


\begin{acknowledgments}
O.D. has received funding from the European Union's Horizon
2020 research and innovation programme under the Marie
Sklodowska-Curie grant agreement No 713694 (MULTIPLY). This
work was partly supported by the Spanish Government through grant
FIS-2015-69617-C2-1-P (J.R.-L.) and the Alexander von Humboldt
Foundation within the Renewed research stay program (E.K.).
\end{acknowledgments}

\begin{appendices}
\section*{Appendix}
\subsection{Convergence rate if $\tau=2M T_n$}

In the long time limit, Eq. \eqref{eq:ngen} reduces to
\begin{widetext}
\begin{eqnarray}
{\bf x}_{nA,B} &  =  &  \frac{1}{2}\left[e^{-\lambda_0 T_n} \sum_{k=1}^{n-1}\frac{(\lambda_0 T_n)^k}{k!}\left(\frac{A_1^k {\bf x}_{(n-k)A}+A_2^k {\bf x}_{(n-k)B}}{1-e^{-\lambda_0 T_n}}\mp \frac{A_1^k {\bf x}_{(n-k)A}-A_2^k {\bf x}_{(n-k)B}}{1+e^{-\lambda_0 T}}\right)\right]\label{eq:ngenlongtime}\,.
\end{eqnarray}
\end{widetext}
Inserting a solution 
$$ {\bf x}_{nA,nB} = A^n {\bf x}_{0A,0B}\,,$$
Eq. \eqref{eq:ngenlongtime}  can be approximated as
\begin{small}
\begin{equation}
{\bf x}_{nA,B}  \approx   \frac{1}{1-e^{-2\lambda_0 T_n}}\left[e^{-2\lambda_0 T_n} M_1 {\bf x}_{nA}+e^{-\lambda_0 T_n} M_2 {\bf x}_{nB}\right]\label{eq:ngenlongtime2}\,,
\end{equation}
\end{small}
with 
$$M_{1,2}=e^{\lambda_0 T_n A_{1,2}A^{-1}}-1\,.$$
After a little algebra, this leads to
\begin{eqnarray}
e^{-2\lambda_0 T_n} M_1 M_2 &  =  & (1-e^{-2\lambda_0 T_n} - e^{-2\lambda_0 T_n} M_1) \nonumber\\
& &  (1-e^{-2\lambda_0 T_n} - e^{-2\lambda_0 T_n} M_2) \label{eq:ngenlongtime3}\\
& = &  1- e^{-2\lambda_0 T_n} - e^{-2\lambda_0 T_n} (M_1+M_2)\nonumber\,.
\end{eqnarray}
Inserting the expressions for $M_1$ and $M_2$, we find
\begin{eqnarray}
e^{-2\lambda_0 T_n} (e^{\lambda_0 T_n A_{1}A^{-1}}-1)(e^{\lambda_0 T_n A_{2}A^{-1}}-1)  =   & \nonumber\\
1- e^{-2\lambda_0 T_n} - e^{-2\lambda_0 T_n} (e^{\lambda_0 T_n A_{1}A^{-1}}+ e^{\lambda_0 T_n A_{2}A^{-1}}-2) &  \nonumber\\
\Rightarrow e^{-2\lambda_0 T_n} e^{\lambda_0 T_n A_1A^{-1}} e^{\lambda_0 T_n A_2 A^{-1}}  =  I  &  \label{eq:ngenlongtime4}\,.
\end{eqnarray}
The average network, Eq. \eqref{eq:Tneven}, solves Eq.  \eqref{eq:ngenlongtime4}.

\subsection{Convergence rate if $\tau=(2M+1) T_n$}
If the matrices $A_1$ and $A_2$ commute, it is possible to find an asymptotic solution of Eq. \eqref{eq:n2gen} in the slow network limit. Evaluating Eq.  \eqref{eq:n2gen} along a transverse eigenvector $x$ of the two coupling matrices, and taking the limit $e^{-\lambda_0 T_n}\rightarrow 0$, we find the simplified form
\begin{eqnarray}
x_{nA,B} & = & 
e^{-\lambda_0 T_n}\left[\sum_{k=0}^{(n-2)/2}\frac{(\lambda_0 T_n)^{2k+1}}{(2k+1)!} \gamma_{2,1} \tilde{\gamma}^{2k} x_{(n-2k-1)A,B}\right.\nonumber\\ 
&  &\left. + \sum_{k=1}^{(n-2)/2}\frac{(\lambda_0 T_n)^{2k}}{(2k)!}\tilde{\gamma}^{2k} x_{(n-2k)B,A}\right] \label{eq:n2gensimple} \,,
\end{eqnarray}
with $\gamma_1$ and $\gamma_2$ the respective eigenvalues of $A_1$ and $A_2$ and  $\tilde{\gamma}=\sqrt{\gamma_1\gamma_2}$. Inserting a solution 
$$x_{nA,nB}=\gamma^n x_{0A,0B}\,,$$
in Eq. \eqref{eq:n2gensimple} we find in the long time limit, 

\begin{eqnarray}
x_{nA,B} & \approx & 
e^{-\lambda_0 T_n}\left[\frac{\gamma_{2,1}}{\tilde{\gamma}}\sinh(\lambda_0 T_n \tilde{\gamma}/\gamma) x_{0A,0B}\right.\nonumber\\
& & \left.+ (\cosh(\lambda_0 T_n \tilde{\gamma}/\gamma)-1)x_{0B,0A}\right] \,.
\end{eqnarray}

Assuming $Re(\gamma/\tilde{\gamma})>0$, in the limit of $e^{-\lambda_0 T_n} \rightarrow 0$ this simplifies to
\begin{equation}
x_{0A,0B} =\frac{1}{2}e^{\lambda_0 T_n(\tilde{\gamma}/\gamma-1)}\left[\frac{\gamma_{1,2}}{\tilde{\gamma}}x_{0A,0B}+ x_{0B,0A}\right] \,,
\end{equation}
which is solved by
\begin{equation}
\gamma =\frac{\tilde{\gamma}}{1+\frac{1}{\lambda_0 T_n}(\ln[\tilde{\gamma}]-\ln[(\gamma_1+\gamma_2)/2])}\,.
\end{equation}

Assuming $Re(\gamma/\tilde{\gamma})<0$, in the limit of $e^{-\lambda_0 T_n} \rightarrow 0$ we find a solution
\begin{equation}
\gamma =\frac{-\tilde{\gamma}}{1+\frac{1}{\lambda_0 T_n}(\ln[\tilde{\gamma}]-\ln[-(\gamma_1+\gamma_2)/2])}\,.
\end{equation}

Thus, one finds the slowest decay rate when choosing the sign of $\tilde{\gamma}$ such that its argument is closest to the argument of $\gamma_1+\gamma_2$.
\end{appendices}

\end{document}